\documentclass[
 reprint,
amsmath,amssymb,
aps,
prab,
floatfix,
showkeys
]{revtex4-2}

\usepackage{graphicx}
\usepackage{dcolumn}
\usepackage{bm}
\usepackage{hyperref}

\usepackage{xcolor}
\usepackage{ulem}
\usepackage{subcaption}
\usepackage{siunitx}
\usepackage[capitalise]{cleveref}
\usepackage{microtype}

\usepackage{tikz}
\usetikzlibrary{shapes,arrows,graphs}
\tikzstyle{decision} = [diamond, draw, fill=blue!20, 
    text width=4.5em, text badly centered, node distance=3cm, inner sep=0pt]
\tikzstyle{block} = [rectangle, draw, fill=blue!20, 
    text width=5em, text centered, rounded corners, minimum height=4em]
\tikzstyle{line} = [draw, -latex']
\tikzstyle{cloud} = [draw, ellipse,fill=red!20, node distance=3cm,
    minimum height=2em]

\newcommand{\fh}[2]{\textcolor{purple}{#1\ifx&#2&\else\ (#2)\fi}}
\newcommand{\fv}[2]{\textcolor{orange}{#1\ifx&#2&\else\ (#2)\fi}}

\newcommand{\Cp}{C_\mathrm{P}}
\newcommand{\Ca}{C_\mathrm{A}}

\newcommand{\Cd}{C_\mathrm{D}}
\newcommand{\Cc}{C_\mathrm{C}}
\newcommand{\Cfd}{C_\mathrm{FD}}
\newcommand{\cma}{c_\mathrm{ma}}
\newcommand{\repourl}{https://gitlab.cern.ch/btp-public/adpt}

\begin{document}

\preprint{APS/123-QED}

\title{Differentiable simulations for particle tracking in accelerators: analysis, benchmarking and optimization}

\author{Francisco Huhn}
 \email{francisco.huhn@cern.ch}
 \affiliation{CERN, Geneva, Switzerland}
\author{Francesco M. Velotti}
 \email{francesco.maria.velotti@cern.ch}
 \affiliation{CERN, Geneva, Switzerland}

\date{\today}

\begin{abstract}
Optimization of beamlines and lattices is a common problem in accelerator physics, which is usually solved with semi-analytical methods and numerical optimization routines.
However, these are usually of the gradient-free or finite-differences type, whose computational cost grows quickly with the number of optimization parameters.
On the other hand, the cost of gradient-based optimization can scale well with the number of parameters, but only if the computation of the gradient is itself efficient (e.g. not via finite differences, which are inefficient).
Recently, there has been an emergence of so-called ``differentiable'' codes that efficiently provide gradients.
Nevertheless, analysis and benchmarking comparisons of these techniques have largely been absent from the literature.

In this work, we develop our own differentiable code\footnote{The code used in this paper can be found \href{\repourl}{here}.}, via auto-differentiation. We analyze and benchmark differentiability against finite differences in two test cases, a space-charge FODO cell from the literature and a realistic future beamline at CERN. The analyses and benchmarking are done both theoretically and empirically.
Finally, we embed the gradient provided by the differentiable code in gradient-based optimization routines and compare with gradient-free methods.
This work offers the first such analysis and benchmarking and thus contributes towards the development of more efficient and performant particle accelerators.
\end{abstract}

\maketitle


\section{Introduction}
\label{sec:intro}

Optimization of beamlines or lattices is a common problem in accelerator physics in which one or a handful of objectives is minimized/maximized, potentially subject to constraints, by varying a number of parameters, such as quadrupole strengths, etc. Common objectives are Twiss functions matching~\cite{mad_matching}, beam size minimization (e.g. \cite{awake}), beam-distribution tailoring~\cite{Remta2025}, etc. 
Numerical optimization algorithms, in contrast with manual or grid search, have been successfully applied to efficiently solve these problems (e.g.~\cite{awake, lattice_optimization}).
%
These numerical routines can be provided by an external package and paired with a accelerator-physics simulation software or be provided by the simulation tool itself.
MAD-X \cite{MAD-X} is among the most widely used simulation packages in accelerator physics and provides both gradient-based, via finite differences, and gradient-free, like Simplex \cite{Nelder1965}, optimization routines.
More recent tools such as Xsuite \cite{Xsuite} provide comparable functionality for lattice tuning. Nevertheless, for tracking-based optimization or for the global design of a lattice (e.g. element count and placement), practitioners predominantly adopt gradient-free methods, as the finite-difference evaluation of derivatives incurs a prohibitive computational cost.
However, gradient-free algorithms suffer from the ``curse of dimensionality''~\cite{Hansen2010}, where the computational cost quickly becomes prohibitive as the number of parameters increases, rendering their application unfeasible in mid- to high-dimensional design spaces\footnote{What is considered mid and high dimensional depends on the problem and the computational resources available.}.

On the other hand, gradient-based algorithms, which make use of the derivative of the objective (and constraints) with respect to the design parameters, in conjunction with the adjoint technique for gradient computation, are computationally efficient -- they have been extensively applied in various fields, from fluid mechanics~\cite{Jameson1988,Magri2019}, where simulations can range from thousands to hundreds of millions of degrees of freedom, to machine learning~\cite{OWID2024}, where models with up to billions of parameters are optimized.
The use of the adjoint technique is critical, as naive computation of the gradient, such as via finite differences, is inefficient, because it suffers from the same curse of dimensionality as gradient-free methods (see \cref{sec:fd}).
In this technique, the adjoint system is obtained and used to propagate adjoint vectors backward in time, which allows for an efficient computation of the gradient (see~\cref{sec:differentiation}). Traditionally, the adjoint system would be derived analytically and then numerically implemented in code. However, this approach has several disadvantages, being:
i) laborious, as deriving the adjoint system analytically can quickly become unwieldy;
ii) error prone, as the mathematical expressions must be manually transposed;
iii) cumbersome to maintain, especially in updates (e.g. when a new feature is implemented).
Alternatively, one can make use of auto-differentiation (see~\cref{sec:autodiff}, which breaks down all computational operations into a graph of elementary operations, over which applying the chain rule becomes straightforward. This approach suffers from none of the aforementioned issues, but comes at the cost of building the graph, i.e. computational overhead, and storing it, i.e. memory cost. It is this method that has enabled and powered the successful training of large models in the field of machine learning, and is provided by the most popular libraries, such as PyTorch~\cite{Ansel2024}, TensorFlow~\cite{Abadi2015}, JAX~\cite{Jax2025}, etc.

In accelerator physics, the application of the adjoint technique (be it explicit or via auto-differentiation) is scarce. While new so-called ``differentiable'' simulation codes have emerged~\cite{GonzalezAguilera2023,Kaiser2024,JuTrack,Deniau2025}, they are still far from mainstream in the community.
More importantly, use-case examples are often of the tutorial class, i.e. too simple, and comprehensive benchmarking and comparisons of the methods -- both qualitative and quantitative -- are largely absent from the literature.
In fact, the only empirical comparison between finite differences and auto-differentiation is found in a passage in~\cite{JuTrack}, which states that auto-differentiation takes twice as long as (central) finite differences to compute the same gradient.
However, a relevant detail absent from the paper, but mentioned in the code repository~\cite{JuTrackRepo} is that only forward auto-differentiation was implemented. Since the problem has seven parameters and one objective, reverse (i.e. adjoint), not forward, auto-differentiation is likely the optimal method (see~\cref{sec:differentiation} for a discussion on forward vs adjoint).
Therefore, to the best of our knowledge, this publication represents the first time-performance benchmarking and comparison between finite differences and auto-differentiation, as well as between gradient-free and gradient-based optimization.

\section{Gradient computation}

Before delving into the details of the computation of the gradient, let us define the problem.
The beamline is composed of $N$ elements. Starting from the initial state $\bm x_0 \in \mathbb R^{N_p \times 6}$ (if using 6 dimensions for each particle), where $N_p$ is the number of particles, we can track the particles by iterating over the elements, i.e. by sequentially applying the (potentially nonlinear) transfer maps of each element
\begin{equation}
    \bm x_n = \bm f_n(\bm x_{n-1}; \bm p_n)
    \label{eq:primal}
\end{equation}
where $\bm f_n$ is the transfer map and $\bm p_n$ the vector of parameters of the $n$-th element. \Cref{eq:primal} is called the primal and is what is commonly meant by ``simulation''.

We wish to minimize the objective
\begin{equation}
    L(\bm x_0, \dots, \bm x_N; \bm p_1, \dots, \bm p_N),
\end{equation}
which depends on the states, $\bm x_n$, and on the parameters, potentially explicitly (e.g. one might wish to add a penalty proportional to magnet strengths), but always implicitly via the states, as $\bm x_n = \bm g_n(\bm x_0; \bm p_1, \dots, \bm p_n)$.
For simplicity, yet, without loss of generality (generality can be achieved via composition), we will assume going forward that only the first element has parameters, $\bm p_1 = \bm p = (p_1 \dots p_m)$, and the objective function only depends on the final state of all the states, i.e. we have $L(\bm x_N; \bm p)$.

\subsection{Finite Differences}
\label{sec:fd}

In finite differences (FD), the gradient is computed by computing each partial derivative via a perturbation of the primal. For example, in the simplest case, forward FD, which approximates derivatives to first order:
\begin{equation}
    \frac{\partial L}{\partial p_i} \approx
        \frac{L(\bm x_N; p_1, \dots, p_i + \epsilon, \dots, p_m) - L(\bm x_N, p_1, \dots, p_m)}{\epsilon}.
\end{equation}
To obtain the whole gradient, i.e. derivatives for all parameters, one must compute $m$ perturbed primals and one unperturbed. In central FD (second order), $2m$ perturbed primals are needed. Generally, denoting the cost of one primal by $\Cp$, the cost of FD is
\begin{equation}
    \Cfd = (a m + b) \Cp,
\end{equation}
where the constants $a$ and $b$ depend on the type of FD (e.g. $a=1, b=1$ for forward, $a=2,b=0$ for central), i.e. the cost of computing the gradient is proportional to the number of parameters and cost of a primal.

\subsection{Differentiation}
\label{sec:differentiation}

Alternatively, let us attempt to compute the gradient via calculus
\begin{equation}
    \frac{dL}{d\bm p} = \frac{\partial L}{\partial \bm p} + \frac{\partial L}{\partial \bm x_N} \frac{\partial \bm x_N}{\partial \bm p}.
\end{equation}
The first term of the RHS is straightforward to compute. Focusing on the second term, we successively apply the chain rule, obtaining
\begin{align}
    \frac{\partial L}{\partial \bm x_N}&\frac{\partial \bm x_N}{\partial \bm p} =  \frac{\partial L}{\partial \bm x_N}
        \left. \frac{\partial \bm f_N}{\partial \bm x} \right|_{\bm x_{N-1}}
        \cdots \;
        \left. \frac{\partial \bm f_1}{\partial \bm p} \right|_{\bm x_0},
    \label{eq:chain}
\end{align}
where the terms $\partial \bm f_n/\partial \bm x$ are the Jacobians of the transfer map of each element.
Denoting the dimension of the system $D$ (e.g. $D = 6N_p$), the first term of \cref{eq:chain} is a $1 \times D$ vector. Each of the Jacobians is equivalent to a $D \times D$ matrix. Finally, the last term is a $D \times m$ matrix.

The RHS of \cref{eq:chain} can be computed from right to left or left to right. 
Right to left is called forward\footnote{Not to be confused with forward FD, which is unrelated.} or tangent mode. In this case, we have an $N-1$-long series of $D \times D$ by $D \times m$ multiplications, followed by one $1 \times D$ by $D \times m$, resulting in a computational cost of
\begin{equation}
    \Cc = \left((N-1) D^2 m + Dm\right) \cma,
    \label{eq:cost_r2l}
\end{equation}
where $\cma$ is the cost a of scalar multiplication-addition.
Left to right is called reverse or adjoint mode. Analogously, we have an $N-1$-long series of $1 \times D$ by $D \times D$ multiplications, followed by one $1 \times D$ by $D \times m$, resulting in a complexity of
\begin{equation}
    \Cc = \left((N-1) D^2 + Dm\right) \cma.
    \label{eq:cost_l2r}
\end{equation}
Clearly, from \cref{eq:cost_r2l,eq:cost_l2r}, adjoint mode is advantageous as $m$ is increased, since $m$ is multiplied only by $D$, in contrast with $(N-1) D^2 + D$ in tangent.
This is the case because we are considering the situation of one objective function and $m$ parameters. If, instead, we had $m$ objective functions and one parameter, the situation would be reversed.
When the number of parameters is greater than the number of objectives, one should prefer the adjoint approach; and tangent, otherwise.

Irrespective of the approach, adjoint versus tangent, to calculate the gradient via \cref{eq:chain}, the states $\bm x_n$ are required (i.e. the primal must be computed), as well as the terms $\partial f_n/\partial \bm x|_{\bm x_{n-1}}$, necessary to differentiate the transfer maps.
Thus, the total cost of the method in terms of the cost of the primal is
\begin{equation}
    \frac{\Ca}{\Cp} = \frac{\Cp + \Cd + \Cc}{\Cp},
    \label{eq:total_cost}
\end{equation}
where $\Cp$, $\Cd$ and $\Cc$ are the costs of the primal, differentiation and the computation of the chained derivative by multiplication-addition of the chain of the partial derivatives in adjoint mode.
The cost of differentiation, $\Cd$, is $O(\Cp)$\footnote{For example, if the primal is $f(x,y) = x y$, then $\partial_{\bm p} f = y \,\partial_{\bm p} x + x \,\partial_{\bm p} y$ is $y$ and $x$, i.e. $\Cd = 2 \Cp$.}, which we write, foregoing any dominated terms, as
\begin{equation}
    \Cd = \kappa_\mathrm{dp} \Cp.
    \label{eq:cost_diff}
\end{equation}
The relative cost of computing the chained derivative, $\Cc/\Cp$, depends on the problem, i.e. the primal, and the approach, tangent or adjoint.
For example, for a full linear primal, $\Cp = O(N D^2)$, whereas, generally, nonlinearities will introduce higher computational complexity.
%
\Cref{eq:cost_l2r} for $\Cc$ and \cref{eq:cost_diff} can be substituted in \cref{eq:total_cost} to obtain the total cost
\begin{equation}
    \frac{\Ca}{\Cp} = 1 + \kappa_\mathrm{dp} + \kappa_\mathrm{sp} + \kappa_\mathrm{mp}m,
    \label{eq:adjoint_cost}
\end{equation}
where $\kappa_\mathrm{sp} = \cma (N-1)D^2/\Cp$ and $\kappa_\mathrm{mp} = \cma D/\Cp$.
%
Therefore, while, strictly speaking, the adjoint's cost scales linearly with the number of parameters (similarly to FD), the linear coefficient, $\kappa_\mathrm{mp}$, is small to negligible.
Additionally, $\kappa_\mathrm{sp} \rightarrow 0$ as the computational complexity of the primal increases.
%
For example, even with a, computationally simple, linear primal, $\kappa_\mathrm{mp} = \cma D/(\cma ND^2) = 1/(ND)$ and $\kappa_\mathrm{sp} = \cma(N-1)/(\cma N) = (N-1)/N$.
%


This highlights the big advantage of the adjoint method: computing the gradient on one parameter costs approximately the same as on any number of parameters\footnote{Practical limitations, like memory, aside.}, which contrasts with FD, where the linear coefficient is $O(1)$.
For $m$ objective functions and one parameter, analysis of the tangent approach, i.e. forward propagation, results in similar expressions.

\subsection{Auto-differentiation}
\label{sec:autodiff}

While one may derive the analytical expressions of $\partial \bm f_n/\partial \bm x_n$ and other terms, as mentioned in \cref{sec:intro}, this is laborious, error prone and hard to maintain. Instead, auto-differentiation (AD) can be used. In AD, complex computations, e.g. the primal, are broken down into a directed acyclic graph of elementary operations, whose derivatives are known, over which the chain rule can be applied successively.
An example of reverse AD (AD-REV) for a thin-sextupole kick is given in \cref{app:autodiff:example} and a cost analysis of the method is given in \cref{app:autodiff:cost}.

Auto-differentiation found great applicability and success in machine learning~\cite{Baydin2018}, with the most popular libraries (e.g. PyTorch~\cite{Ansel2024}, TensorFlow~\cite{Abadi2015}, JAX~\cite{Jax2025}) providing it as an out-of-the-box feature. In the majority of cases, very little setup is needed -- one programs the primal as a function and then calls the auto-differentiation function (tangent of adjoint) on it. Moreover, these libraries provide GPU support, enabling the study of more computationally-demanding cases.
While PyTorch has been framework of choice for new differentiable codes~\cite{GonzalezAguilera2023,Kaiser2024}, we implement ours\footnote{The code used in this paper can be found \href{\repourl}{here}.} on top of JAX, which is designed for high-performance numerical computing and provides a NumPy-style API.

\subsection{Computational time vs complexity}
\label{sec:time_vs_complexity}

While it should be clear that above a certain number of parameters or primal computational complexity, (auto-)differentiation should be advantageous with respect to finite differences, it is not clear what this number is. Analyzing the expressions for the computational cost of each method would suggest it is small.
However, the situation is complicated by parallelization. For example, the different perturbed primals of finite differences can be straightforwardly run in parallel. Doing the same for the adjoint is significantly harder.
This is where the distinction between computational time and complexity becomes important. In the end, practitioners are interested in shorter computational times, not necessarily complexity. Thus, it may turn out that, despite having higher complexity, finite differences may be faster than the adjoint. Hence the necessity for the kind of benchmarking and comparison done in this work.
It follows thus that the relationship between computation time and complexity is strongly dependent on the interplay between the problem (system size, density, etc.), the method/numerics (FD, AD and their implementation) and the machine (e.g. CPU, GPU, memory, etc.).
Our tests were run on a machine with Intel Xeon Silver 4110 CPU, \SI{16}{\giga\byte} RAM, Tesla V100 with \SI{32}{\giga\byte} VRAM, which we believe is representative of those available to accelerator-physics practitioners.


\section{Problems and Results}

\subsection{FODO with space charge}
\label{sec:jutrack}

\subsubsection{Single cell}
\label{sec:jutrack:single}

This case is taken from~\cite{JuTrack}. It is a 6D tracking problem of 5000 particles over a FODO cell with space-charge effects\footnote{Details can be found in the reference}.
Following the reference, each of the drifts is sliced into segments of 0.05 m and the quadrupoles are sliced into segments of 0.025 m. 
The drift segments are integrated via drift-kick-drift, with the kick due to space-charge effects.
The quadrupole segments are similarly halved, with space-charge kick applied between the halves, with each half being further subdivided into 20 sub-segments, with each sub-segment being integrated via the 7-step Yoshida integrator (fourth order).
The objective functions are the horizontal and vertical emittances. There are seven parameters, five lengths (one per each of the three drifts and one per each of the two quadrupoles) plus two strengths (one per quadrupole).

\begin{figure}[htbp]
    \centering
    \includegraphics[width=\linewidth]{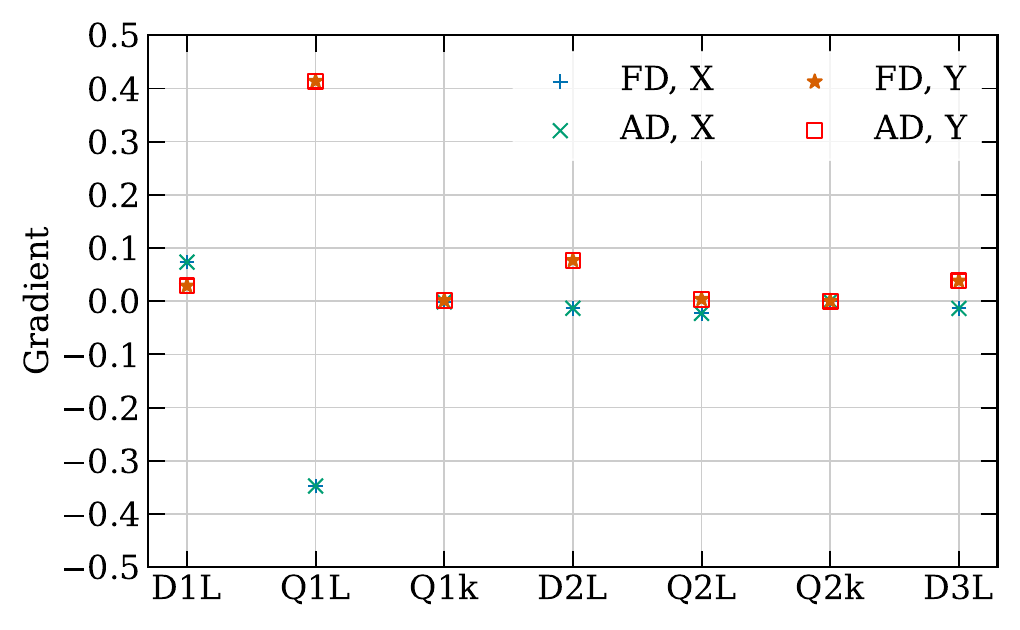}
    \caption{Gradients of horizontal ($X$) and vertical ($Y$) emittances, via auto-differentiation (AD) and finite differences (FD).}
    \label{fig:jutrack:sensitivities}
\end{figure}

\begin{figure*}[tb]
    \centering
    \begin{subfigure}[b]{0.475\linewidth}
        \centering
        \includegraphics[width=\linewidth]{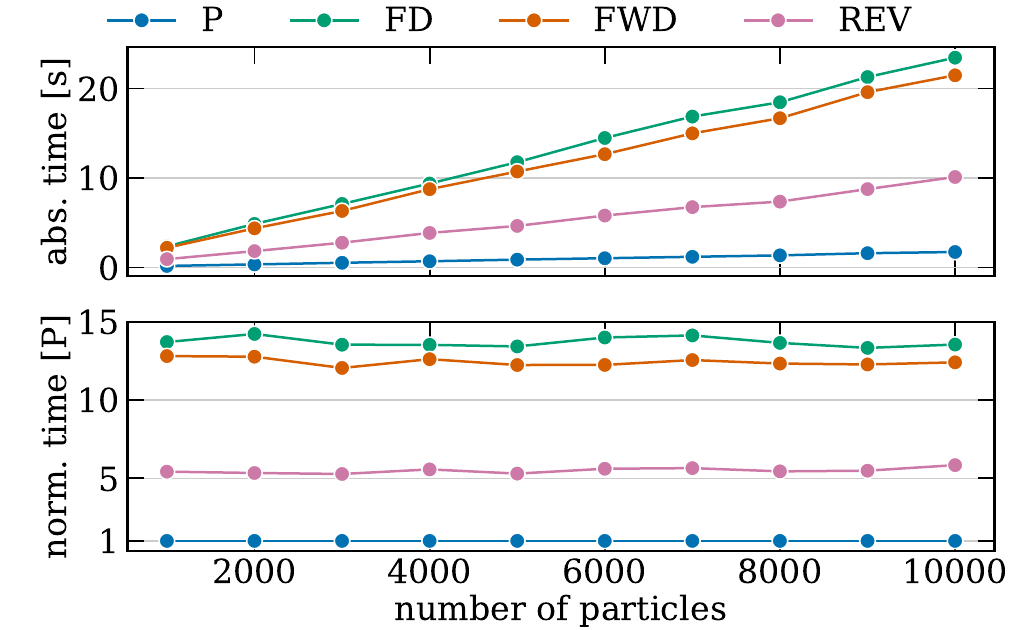}
        \caption{CPU}
        \label{fig:jutrack:cell_times:cpu}
    \end{subfigure}
    \hspace{0.025\linewidth}
    \begin{subfigure}[b]{0.475\linewidth}
        \centering
        \includegraphics[width=\linewidth]{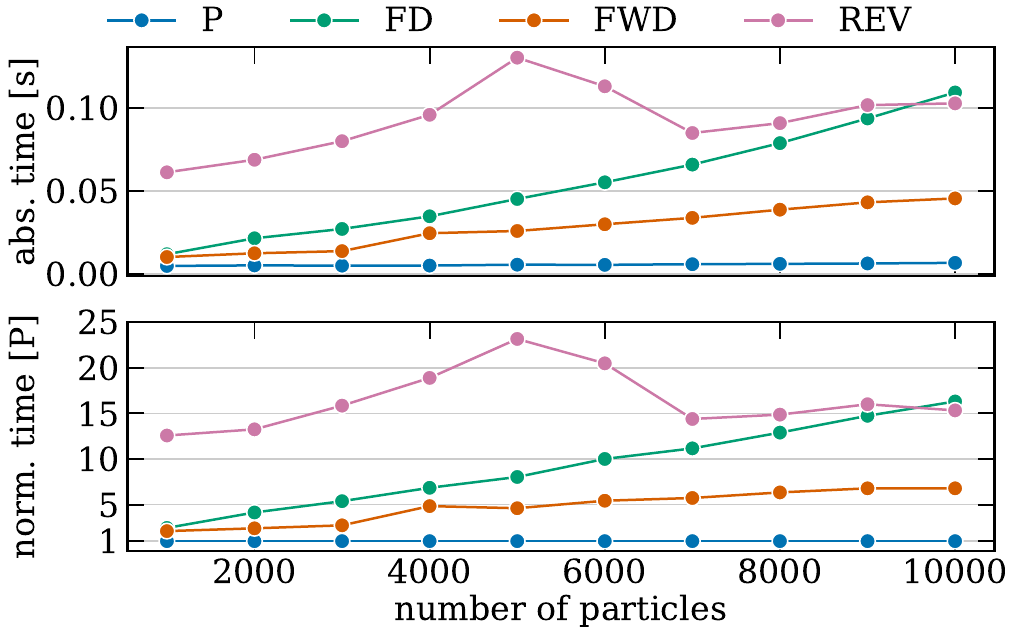}
        \caption{GPU}
        \label{fig:jutrack:cell_times:gpu}
    \end{subfigure}
    \caption{Execution times of primal (P), finite differences (FD), adjoint auto-differentiation (REV), tangent auto-differentiation (FWD). Top panels: absolute time; bottom panels: normalized by primal time.}
    \label{fig:jutrack:cell_times}
\end{figure*}

We compare the values of the gradients obtained via AD (adjoint) and FD (central FD with step of $10^{-6}$ \cite{JuTrack}).
\Cref{fig:jutrack:sensitivities} shows great agreement between the two methods and the values are also very similar\footnote{The reason for the slight discrepancies is unknown.} to those found in~\cite{JuTrack}.
\Cref{fig:jutrack:cell_times} shows the execution time for each method FD, AD-FWD, AD-REV, as well as the primal, on both CPU and GPU.
%
On CPU, as the computational complexity (due to the number of particles) increases, the computational times increase. However, normalizing by the cost of a primal (i.e. of one simulation), we can see that AD-REV costs the equivalent of 5 primals, while FD costs close to 3 times more. On the other hand, AD-FWD costs only slightly less than FD, which is not unsurprising given the low number of parameters. Notice, however, that this is in stark contrast with the findings in~\cite{JuTrack} that FD is twice as fast as AD-FWD.
On GPU, however, there is a surprising reversal: AD-REV becomes the least-performing method for the majority of cases ($N_p \leq 9000$) and is only marginally faster than FD at $N_p=10000$. Nevertheless, auto-differentiation still comes ahead with FWD. The reason for these, at first, surprising, results, is alluded to in \cref{sec:time_vs_complexity}. GPUs possess great parallel processing power, which lends itself greatly to FD and AD-FWD, as both can take advantage of this capability due to their embarrassingly-parallel nature, and not so much to AD-REV.
Despite needing more operations, in FD and AD-FWD, these can be executed in parallel. An analysis of time versus complexity in the following section (\cref{fig:jutrack:time_vs_flops}) effectively demonstrates this.
This is an advantage that should subside with increasing problem complexity -- indeed, at $N_p = 10000$, we already observe a reversal in the relative performance of FD vs AD-REV.

\subsubsection{Multiple cells}
\label{sec:jutrack:multiple}

\begin{figure}[h]
    \centering
    \includegraphics[width=\linewidth]{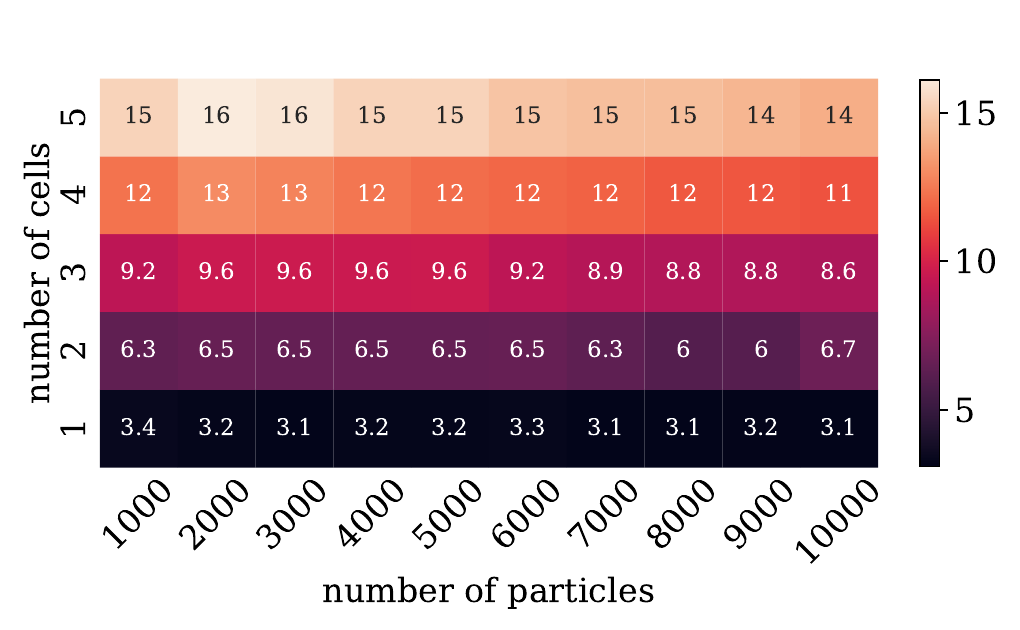}
    \caption{Ratio of execution times: FD over AD-REV -- CPU.}
    \label{fig:jutrack:fd_over_ad_cpu}
\end{figure}

\begin{figure*}[tb]
    \centering
    \begin{subfigure}[b]{0.475\linewidth}
        \centering
        \includegraphics[width=\linewidth]{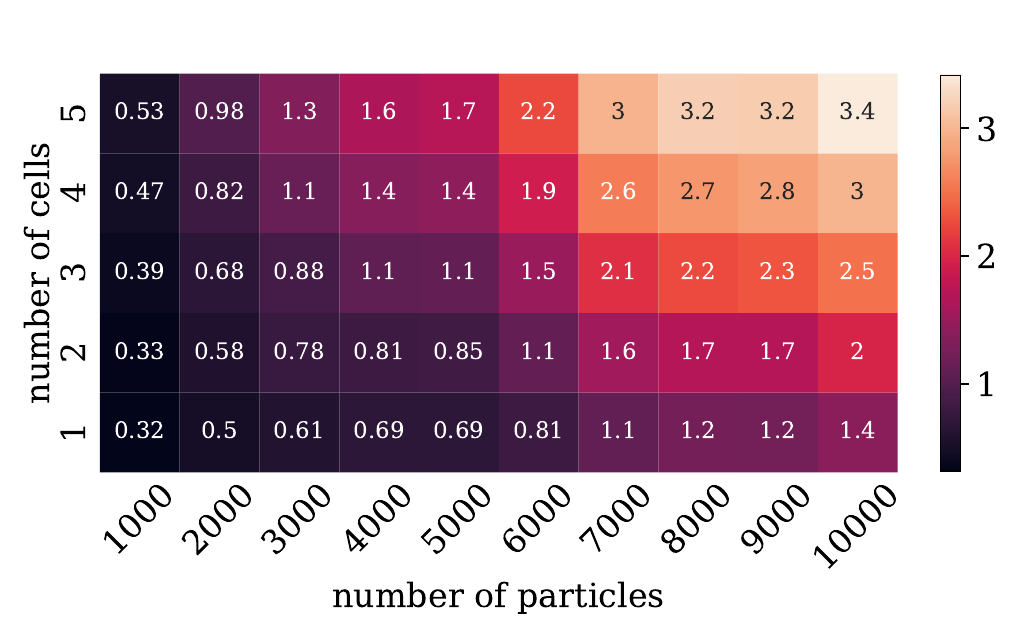}
        \caption{AD-REV}
        \label{fig:jutrack:fd_over_adrev_gpu}
    \end{subfigure}
    \hspace{0.025\linewidth}
    \begin{subfigure}[b]{0.475\linewidth}
        \centering
        \includegraphics[width=\linewidth]{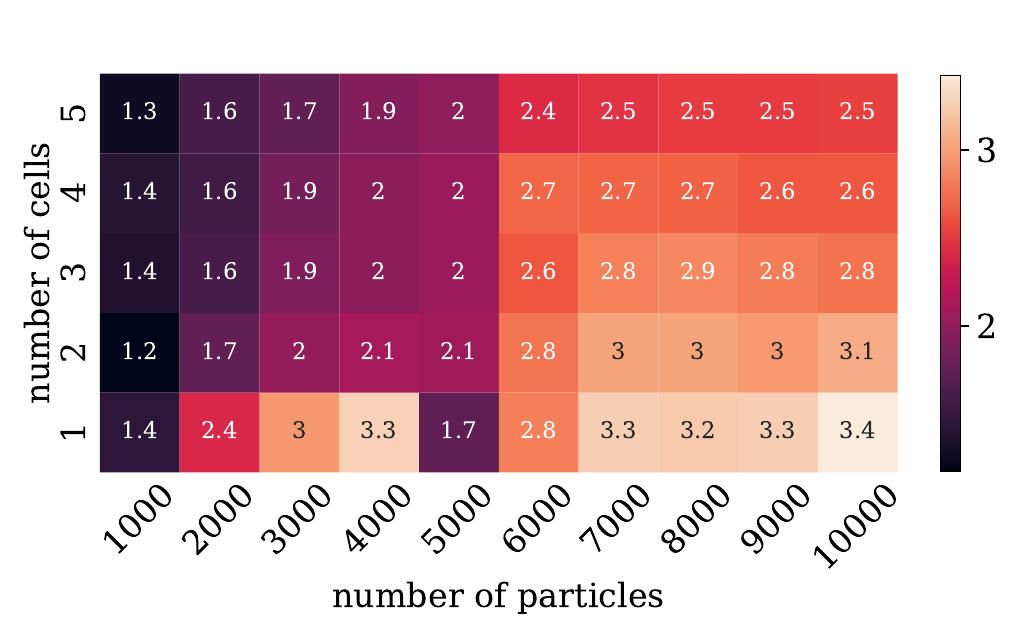}
        \caption{AD-FWD}
        \label{fig:jutrack:fd_over_adfwd_gpu}
    \end{subfigure}
    \caption{Ratio of execution times: FD over AD -- GPU.}
    \label{fig:jutrack:fd_over_ad}
\end{figure*}

\begin{figure}[htbp]
    \centering
    \includegraphics[width=\linewidth]{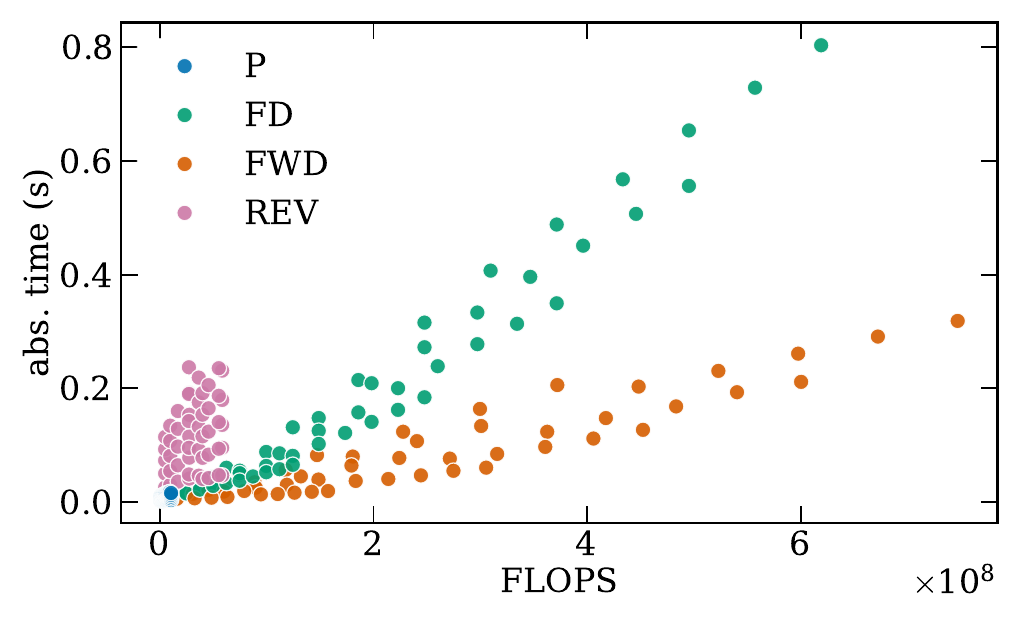}
    \caption{Execution time versus number of operations}
    \label{fig:jutrack:time_vs_flops}
\end{figure}

We slightly modify the FODO cell of the previous section. To slightly reduce the computational cost of each cell, we reduce the number of integration steps of a half slice from 20 to 5. To avoid two consecutive drifts, we double the length of the first drift and remove the last drift. 
Taking this modified cell, we make a beamline of up to 5 cells.
Thus, we not only add computational complexity to the problem, similarly to adding more particles, but also increase the number of parameters. With this increase, the performance gap between FD and AD should become larger (see \cref{sec:differentiation}).
The results for CPU (\cref{fig:jutrack:fd_over_ad_cpu}) show that, as in the previous section, AD-REV is 3 times faster than FD for 1 cell, but for 5 cells it is 15 times faster. This is not surprising, since the computational cost of AD-REV is practically constant in primal terms, whereas that of FD increases linearly with the number of parameters.
On the other hand, if the code is executed on GPU (\cref{fig:jutrack:fd_over_adrev_gpu}), we can see that FD is faster than AD-REV for cases of low number of cells and low number of particles. Once again, this shows the potential of FD to take advantage of parallelization in ``small'' problems. However, in these cases, AD-FWD is even faster than FD (\cref{fig:jutrack:fd_over_adfwd_gpu}).
The computational time vs complexity distinction can be clearly seen in \cref{fig:jutrack:time_vs_flops}, where time is plotted versus an estimate of the number of operations\footnote{This estimate is obtained via the \texttt{func.lower(x).compile().cost\_analysis()} method in JAX.}. For the same number of operations, the AD-REV takes longer than FD and AD-FWD.

In essence, for the cases analyzed, AD-REV is generally faster than FD, both on CPU and GPU. When solving small cases on GPU, AD-REV may be slower than FD, but AD-FWD is still faster than FD, which means that auto-differentiation is more efficient than finite differences in all cases tested.

\subsection{Realistic beamline}
\label{sec:awake}

To obtain a more realistic estimate of each method’s performance, we have chosen a high-energy beamline so that collective effects can be safely neglected\footnote{Coherent synchrotron radiation is the only effect that could be considered but for the purpose of this paper, it was considered negligible.}, incorporating both sextupole and octupole magnets. This is the 150~MeV electron transfer line now being designed for the next phase of the CERN AWAKE experiment; it is designed to deliver micrometer-scale transverse beam sizes at the plasma injection point for subsequent acceleration~\cite{awake}.

\begin{figure*}[tb]
    \centering
    \includegraphics[width=\linewidth]{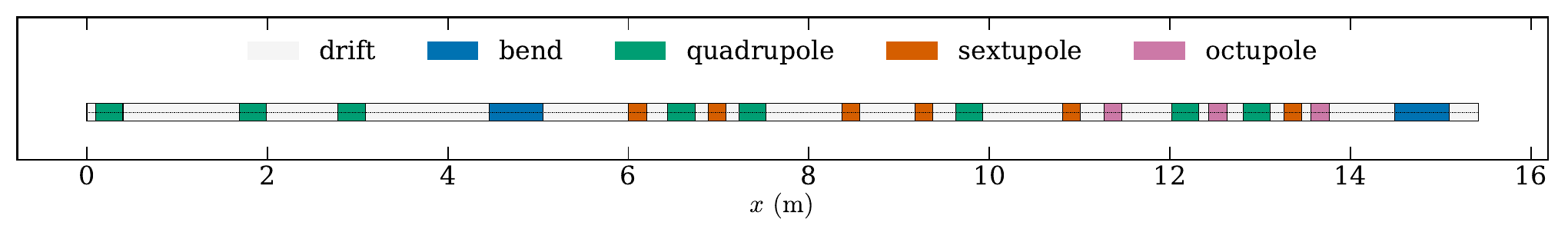}
    \caption{AWAKE line}
    \label{fig:awake:line}
\end{figure*}

Unlike the problem in \cref{sec:jutrack}, there are no collective effects. In this case, far more typical, each particle evolves independently -- the joint system (i.e. system of all particles) is a block-diagonal system, each block corresponding to a particle.
The beamline (\cref{fig:awake:line}) is composed of: drifts, sector bends, quadrupoles, sextupoles and octupoles, for a total of 39 elements, resulting in 58 parameters (\cref{tbl:awake})\footnote{The order of the elements and the values of the parameters of the initial design can be found in the project's \href{\repourl}{repository}.}. With exception of sector bends, all magnets are modeled as drift-kick-drift.
For simplicity, we do not implement fringe fields or particle losses.
\begin{table}[h]
    \centering
    \begin{tabular}{llrr}
        \textbf{Element Type} & \textbf{Parameters} & \textbf{NoE}\footnote{Number of Elements} & \textbf{NoP}\footnote{Number of Parameters} \\ \hline
        drift       & length           & 20 & 20 \\
        sector bend & length, angle    &  2 &  4 \\
        quadrupole  & length, strength &  8 & 16 \\
        sextupole   & length, strength &  6 & 12 \\
        octupole    & length, strength &  3 &  6 \\ \hline
        \multicolumn{2}{c}{\textbf{Total}} &  39 & 58
    \end{tabular}
    \caption{Elements of the AWAKE line.}
    \label{tbl:awake}
\end{table}

\begin{figure}[htbp]
    \centering
    \includegraphics[width=\linewidth]{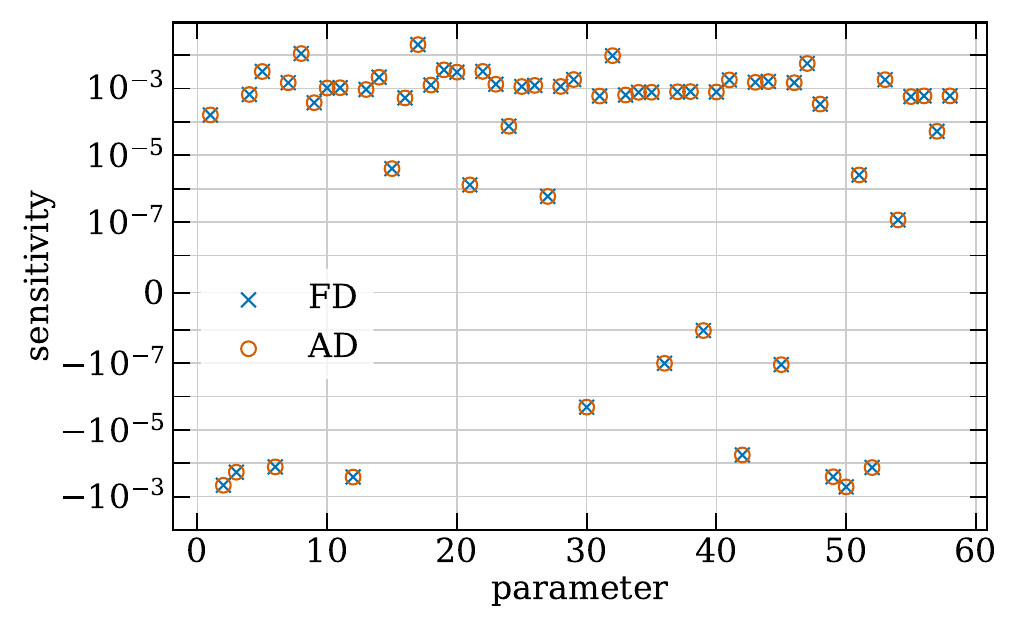}
    \caption{Sensitivity of the objective function (\cref{eq:awake_objective}) with respect to beamline parameters, via auto-differentiation (AD) and finite differences (FD).}
    \label{fig:awake:sensitivities}
\end{figure}

The objective is to minimize
\begin{equation}
    \frac{1}{N_p}\sum_{i=1}^{N_p} (x_i^2 + y_i^2),
    \label{eq:awake_objective}
\end{equation}
i.e. the mean distance from the center.
\Cref{fig:awake:sensitivities} shows that the sensitivities of \cref{eq:awake_objective} with respect to beamline parameters obtained via auto-differentiation match those computed using (central) finite differences.

\begin{figure*}[htbp]
    \centering
    \begin{subfigure}[b]{0.475\linewidth}
        \centering
        \includegraphics[width=\linewidth]{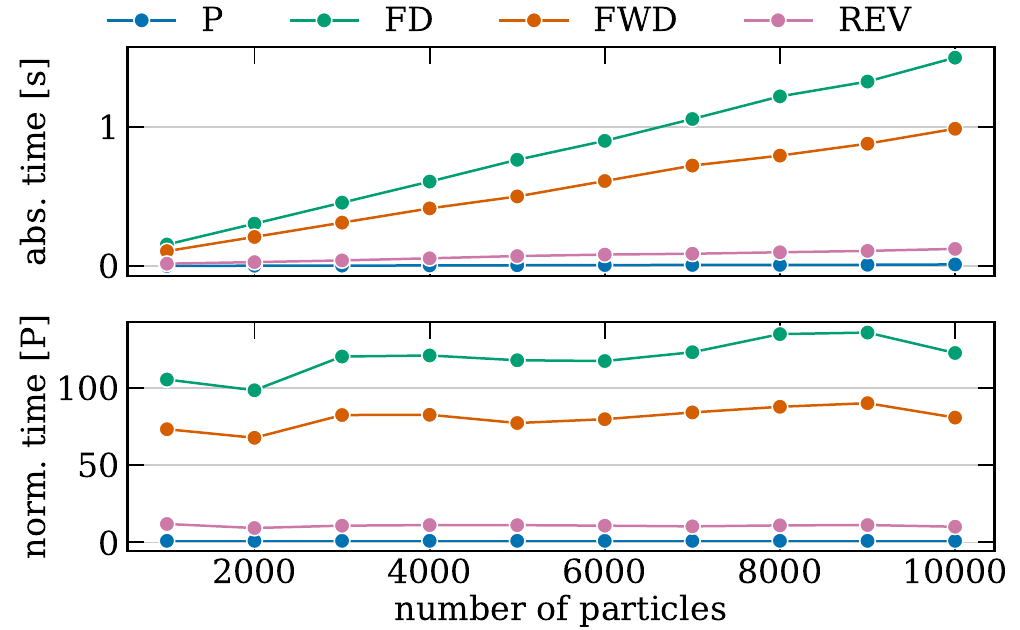}
        \caption{CPU}
        \label{fig:awake:times:cpu}
    \end{subfigure}
    \hspace{0.025\linewidth}
    \begin{subfigure}[b]{0.475\linewidth}
        \centering
        \includegraphics[width=\linewidth]{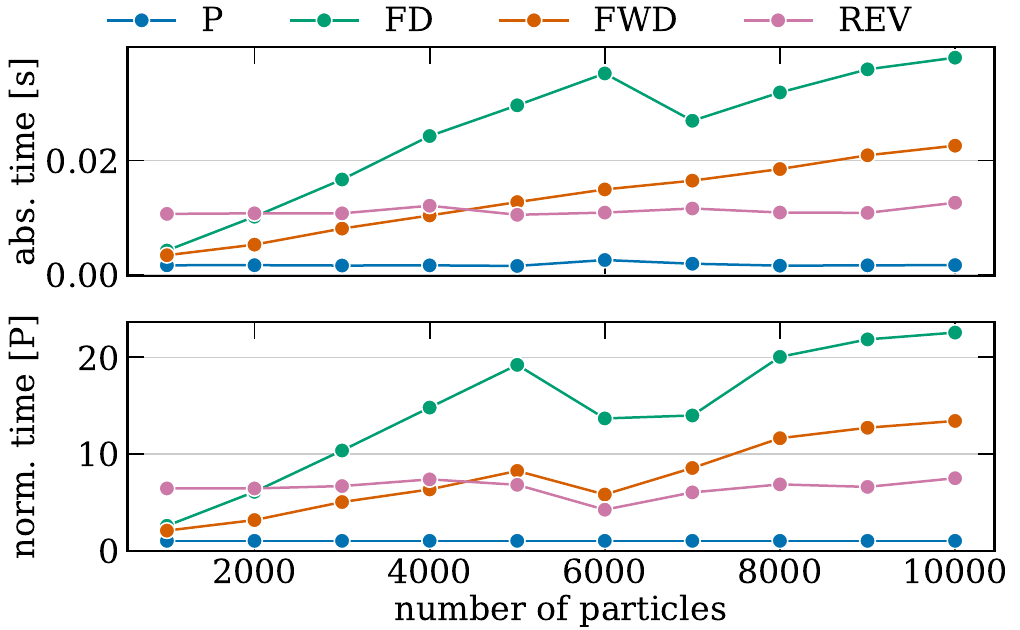}
        \caption{GPU}
        \label{fig:awake:times:gpu}
    \end{subfigure}
    \caption{AWAKE line -- execution times.}
    \label{fig:awake:times}
\end{figure*}

Having validated the numerical values, we now compare the time efficiencies of these methods.
\Cref{fig:awake:times} shows the execution times, absolute and normalized by the time to run one simulation, i.e. one primal.
On CPU, AD-REV is one order of magnitude faster than FD. AD-FWD is also significantly slower than AD-REV, but still faster than FD.
On GPU, AD-FWD is the fastest method for $N_p \leq 4000$ particles, above which it is AD-REV. On the other hand, FD is always slower than either AD-FWD or AD-REV and is the slowest for $N_p \geq 3000$. The speedup from using auto-differentiation ranges from marginal at $N_p=1000$ to a factor of 3 at the upper end of range of particles tested.
%
We remark that the lower computational complexity of the primal (no collective effects and lower number of slices) with respect to the cases in \cref{sec:jutrack}\footnote{Confirmed by the primals' execution times (see \cref{fig:jutrack:cell_times:cpu} vs \cref{fig:awake:times:cpu})} is a factor that benefits FD, relative to AD (see \cref{sec:differentiation}).
Nevertheless, we observe the opposite: the overperformance of AD over FD increases. The reason is that the number of parameters increases from $\leq 30$ to 59.
This comparison highlights the two positive factors of overperformance of AD with respect to FD: computational complexity of the primal and number of parameters.

\subsubsection{Optimization}
\label{sec:awake:optim}

\begin{figure}[htbp]
    \centering
    \includegraphics[width=\linewidth]{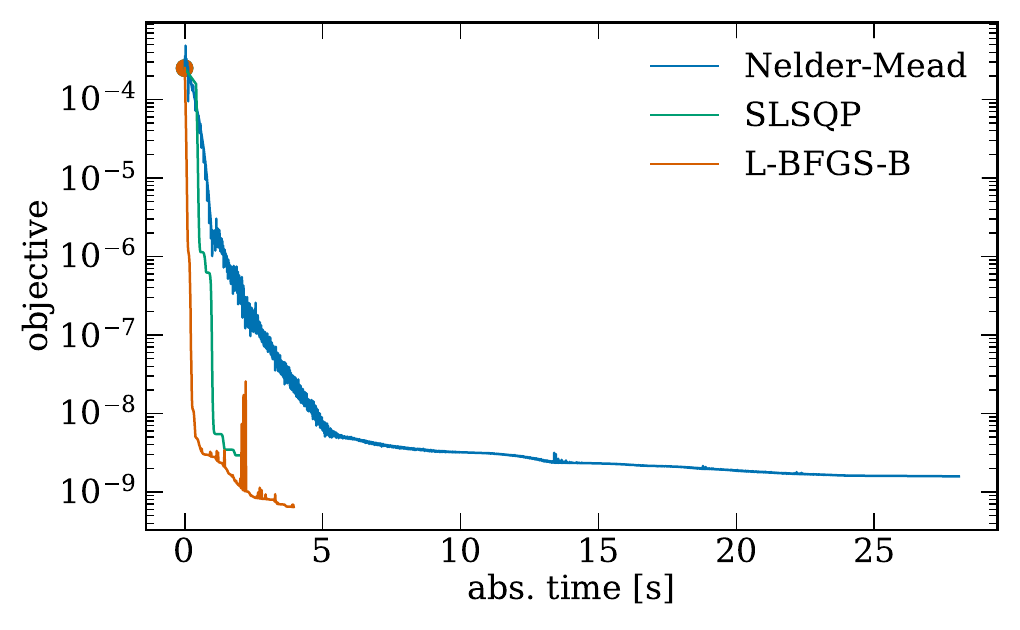}
    \caption{Objective versus time for each of the methods.}
    \label{fig:awake:obj_vs_time}
\end{figure}

\begin{figure*}[htbp]
    \centering
    \includegraphics[width=\linewidth]{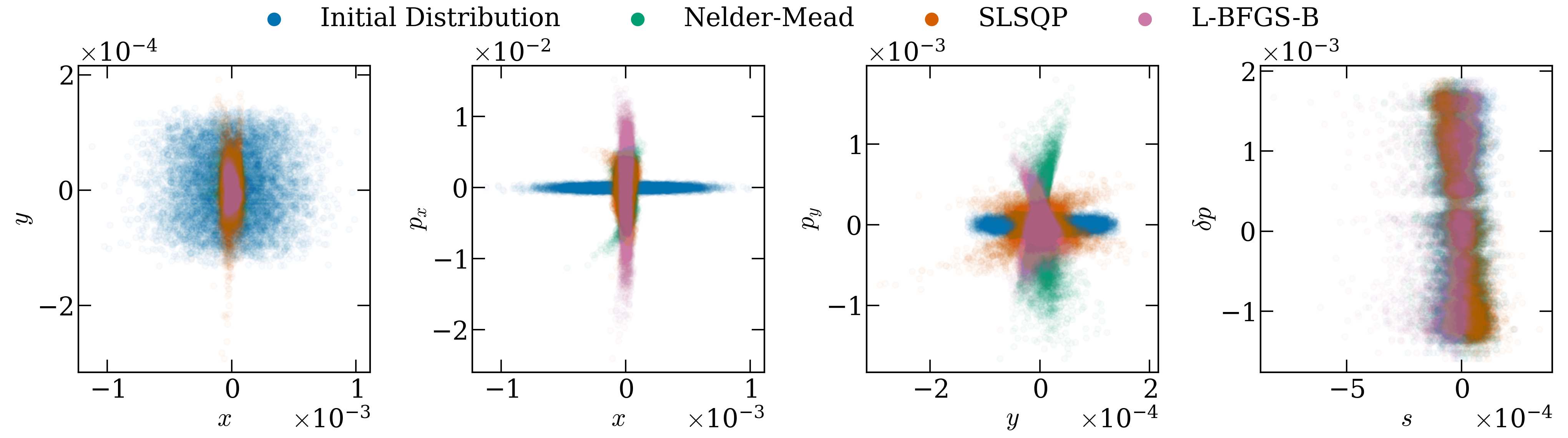}
    \caption{Initial distribution and final distributions for each of the methods. The final distribution of the initial design is not shown because it is much larger than the plotting ranges.}
    \label{fig:awake:distributions}
\end{figure*}

Even if AD is more efficient than FD at computing the gradient, it does not necessarily follow that gradient-based optimization with AD is more efficient than gradient-free methods.
%
Therefore, we compare gradient-based methods, SLSQP \cite{Kraft1994} and L-BFGS-B \cite{Zhu1997}, with one of the most common and performant gradient-free methods, Nelder-Mead \cite{Nelder1965} (it dominated the gradient-free methods we tested).
For a balance between a more realistic scenario (e.g. no nonsensical negative-length drifts) with simplicity, we bound the parameters to $\pm 50\%$ of their initial values.
\Cref{fig:awake:obj_vs_time} shows the evolution of each of the optimizations over time. It shows that both gradient-based methods achieve their optima faster than the gradient-free method and that their optima are better (lower objective function).
\Cref{fig:awake:distributions} shows the initial distribution and each of the three final distributions, showing the noticeable improvement in the objective.

\section{Conclusion}
\label{sec:conclusion}

We analyzed, both theoretically and empirically, the efficiency of finite differences versus auto-differentiation, concluding that AD-REV is almost always faster than FD.
The speedup can range from marginal to orders of magnitude and depends on the physical problem and the computational machine. It is more significant on CPU than on GPU, owing to the latter's parallel processing power, of which FD can better take advantage.
Notwithstanding, even in cases where FD is faster than AD-REV, it is slower than AD-FWD.
Moreover, we embedded the sensitivities calculated via AD-REV in a gradient-based optimizer to solve a parameter-bounded problem, and compared that with a common gradient-free method, finding that the gradient-based approach is faster and reaches a better minimum.

This work shows that differentiable simulations present significant advantages in beam-line optimization and have the potential to improve accelerator design in the search for higher and more robust performance.


\appendix

\section{Auto-differentiation}

\subsection{Example}
\label{app:autodiff:example}

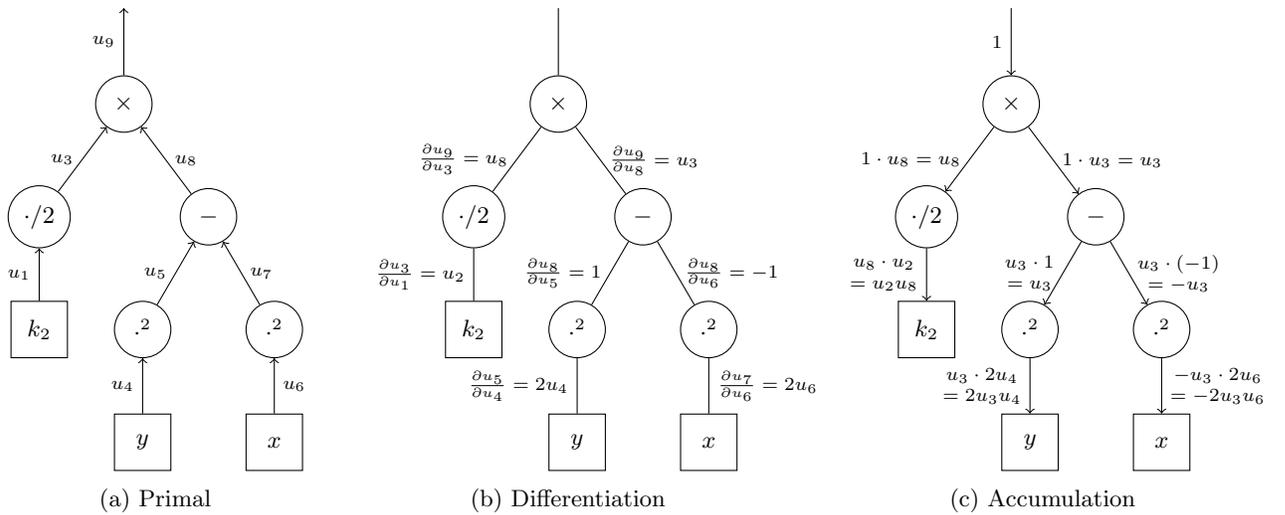
\begin{figure*}[tb]
  \centering
  \begin{subfigure}[b]{0.3\linewidth}
    \begin{tikzpicture}[
        result/.style={diamond},
        operation/.style={circle, draw, minimum size=0.75cm},
        leaf/.style={rectangle, draw, minimum size=0.75cm},
        every node/.style={align=center},
        level 1/.style={sibling distance=2cm}, 
        level 2/.style={sibling distance=2.25cm}, 
        level 3/.style={sibling distance=1.75cm}, 
        level distance=1.5cm,
        edge from parent/.style={<-,draw,font=\scriptsize}
    ]
    \node[result] (root) {}
        child { node[operation] {$\times$}
            child {
                node[operation] {$\cdot/2$}
                child {
                    node[leaf] {$k_2$}
                    edge from parent node[left]{$u_1$}
                }
                edge from parent node[left] {$u_3$}
            }
            child {
                node[operation] {$-$}
                child {
                    node[operation] {$\cdot^2$}
                    child {
                        node[leaf] {$y$}
                        edge from parent node[left] {$u_4$}
                    }
                    edge from parent node[left] {$u_5$}
                }
                child {
                    node[operation] {$\cdot^2$}
                    child {
                        node[leaf] {$x$}
                        edge from parent node[right] {$u_6$}
                    }
                    edge from parent node[right] {$u_7$}
                }
                edge from parent node[right] {$u_8$}
            }
            edge from parent node[left] {$u_9$}
        };
    \end{tikzpicture}
    \caption{Primal}
    \label{fig:ad_dag:primal}
  \end{subfigure}
  %
  %
  \begin{subfigure}[b]{0.3\linewidth}
    \begin{tikzpicture}[
        result/.style={diamond},
        operation/.style={circle, draw, minimum size=0.75cm},
        leaf/.style={rectangle, draw, minimum size=0.75cm},
        every node/.style={align=center},
        level 1/.style={sibling distance=2cm}, 
        level 2/.style={sibling distance=2.25cm}, 
        level 3/.style={sibling distance=1.75cm}, 
        level distance=1.5cm,
        edge from parent/.style={-,draw,font=\scriptsize}
    ]
    \node[result] (root) {}
        child { node[operation] {$\times$}
            child {
                node[operation] {$\cdot/2$}
                child {
                    node[leaf] {$k_2$}
                    edge from parent node[left]{$\frac{\partial u_3}{\partial u_1}=u_2$}
                }
                edge from parent node[left] {$\frac{\partial u_9}{\partial u_3}=u_8$}
            }
            child {
                node[operation] {$-$}
                child {
                    node[operation] {$\cdot^2$}
                    child {
                        node[leaf] {$y$}
                        edge from parent node[left] {$\frac{\partial u_5}{\partial u_4}=2 u_4$}
                    }
                    edge from parent node[left] {$\frac{\partial u_8}{\partial u_5}=1$}
                }
                child {
                    node[operation] {$\cdot^2$}
                    child {
                        node[leaf] {$x$}
                        edge from parent node[right] {$\frac{\partial u_7}{\partial u_6}=2 u_6$}
                    }
                    edge from parent node[right] {$\frac{\partial u_8}{\partial u_6}=-1$}
                }
                edge from parent node[right] {$\frac{\partial u_9}{\partial u_8}=u_3$}
            }
        };
    \end{tikzpicture}
    \caption{Differentiation}
    \label{fig:ad_dag:diff}
  \end{subfigure}
  \hspace*{0.04\linewidth}
  \begin{subfigure}[b]{0.3\linewidth}
    \begin{tikzpicture}[
        result/.style={diamond},
        operation/.style={circle, draw, minimum size=0.75cm},
        leaf/.style={rectangle, draw, minimum size=0.75cm},
        every node/.style={align=center},
        level 1/.style={sibling distance=2cm}, 
        level 2/.style={sibling distance=2.25cm}, 
        level 3/.style={sibling distance=1.75cm}, 
        level distance=1.5cm,
        edge from parent/.style={->,draw,font=\scriptsize}
    ]
    \node[result] (root) {}
        child { node[operation] {$\times$}
            child {
                node[operation] {$\cdot / 2$}
                child {
                    node[leaf] {$k_2$}
                    edge from parent node[left]{$u_8\cdot u_2$\\$=u_2 u_8$}
                }
                edge from parent node[left] {$1\cdot u_8=u_8$}
            }
            child {
                node[operation] {$-$}
                child {
                    node[operation] {$\cdot^2$}
                    child {
                        node[leaf] {$y$}
                        edge from parent node[left] {$u_3\cdot2 u_4$\\$=2u_3u_4$}
                    }
                    edge from parent node[left] {$u_3\cdot1$\\$=u_3$}
                }
                child {
                    node[operation] {$\cdot^2$}
                    child {
                        node[leaf] {$x$}
                        edge from parent node[right] {$-u_3\cdot 2 u_6$\\ $=-2u_3u_6$}
                    }
                    edge from parent node[right] {$u_3\cdot(-1)$\\$=-u_3$}
                }
                edge from parent node[right] {$1\cdot u_3=u_3$}
            }
            edge from parent node[left] {$1$}
        };
    \end{tikzpicture}
    \caption{Accumulation}
    \label{fig:ad_dag:acc}
  \end{subfigure}

  \caption{Reverse auto-differentiation of the horizontal kick of a thin sextupole, $\frac{k_2}{2}(y^2-x^2)$. For simplicity, $u_2=2$, the second argument in $k_2/2$, has been merged into the operation node.}
  \label{fig:ad_dag}
\end{figure*}

As explained in \cref{sec:autodiff}, auto-differentiation represents the numerous computational operations of a computationally-complex function as directed acyclic graph (DAG) of, usually unary or binary, elementary operations.
\Cref{fig:ad_dag} shows such a graph for the horizontal kick of a thin sextupole, which is given by the expression
\begin{equation}
    \Delta x' = \frac{k_2}{2}(y^2-x^2),
\end{equation}
where $k_2$ is the sextupole strength, and $x$ and $y$ are the horizontal and vertical positions.
From left to right and in depth-first order, the expression can be broken down into:
\begin{align}
    &u_1 = k_2 \\
    &u_2 = 2 \\
    &u_3 = u_1/u_2 \\
    &u_4 = y \\
    &u_5 = u_4^2 \\
    &u_6 = x \\
    &u_7 = u_6^2 \\
    &u_8 = u_5 - u_7 \\
    &u_9 = u_3 \times u_8 \\
    &\Delta x' = u_9
\end{align}
This is shown in \cref{fig:ad_dag:primal}. With given values for $k_2$, $x$ and $y$, the graph can be traversed and $\Delta x'$ can be obtained at $u_9$.
Then, starting from the top-most node, i.e. the quantity we wish to differentiate, reversed traversal is executed. At every node with result $u_i$, the partial derivative is computed with respect to each argument (i.e. for each child edge), a value that is then multiplied by the chained derivative to produce the chained derivative of the argument node, i.e. $\partial u_9/\partial u_i$.
For example, the partial derivative of $u_5$ with respect to its argument, $u_4$, is $\partial u_5 / \partial u_4 = 2 u_4$, which is then multiplied by the chained derivative,
\begin{equation}
    u_3 \cdot 2 u_4 = \left(\frac{\partial u_9}{\partial u_8} \frac{\partial u_8}{\partial u_5} \right) \cdot \frac{\partial u_5}{\partial u_4} = \frac{\partial u_9}{\partial u_4}.
    \label{eq:sextupole_kick}
\end{equation}
\Cref{fig:ad_dag:diff,fig:ad_dag:acc} show these steps. The steps are shown in separate figures for clarity. They are applied in succession at each edge of the graph.
Finally, at the leaves, the expressions for the chained gradients are
\begin{equation}
    \begin{pmatrix}
        \frac{\partial u_9}{\partial k_2} \\
        \frac{\partial u_9}{\partial x} \\
        \frac{\partial u_9}{\partial y}
    \end{pmatrix}
    =
    \begin{pmatrix}
        u_2 u_8  \\
        - u_3 u_6 \\
         u_3 u_4
    \end{pmatrix}
    =
    \begin{pmatrix}
        \frac{1}{2}(y^2-x^2) \\
        -k_2 x \\
        k_2 y
    \end{pmatrix},
\end{equation}
which match the analytical differentiation of \cref{eq:sextupole_kick}.
%
%

\subsection{Cost analysis}
\label{app:autodiff:cost}

The AD framework also presents an alternative way to derive the computational cost of differentiation. Here, as in most computational machines, we consider a DAG of unary or binary operations, although the same could be done for any arity.
The first step is to compute the primal via a forward pass, which costs
\begin{equation}
    \Cp = \sum_{i=1}^n c_\mathrm{p}^{(i)},
\end{equation}
where $c_\mathrm{p}^{(i)}$ is the cost of the operation of node $i$ and $n$ is the number of non-leaf nodes.
%
%
The second step is backpropagation, where, for each non-leaf node, one must compute the derivative with respect to each of its children, multiply by the chained derivative and accumulate in the child node.
%
%
The derivatives cost
\begin{align}
    \Cd
    &= \sum_{i=1}^n \left( c_\mathrm{d}^{(i,1)} + c_\mathrm{d}^{(i,2)} \right) \\
    &= \sum_{i=1}^n \frac{c_\mathrm{d}^{(i,1)} + c_\mathrm{d}^{(i,2)}}{c_\mathrm{p}^{(i)}} c_\mathrm{p}^{(i)}\\
    &= \sum_{i=1}^n k_\mathrm{dp}^{(i)} c_\mathrm{p}^{(i)} \\
    &= \kappa_\mathrm{dp} \sum_{i=1}^n c_\mathrm{p}^{(i)} \\
    &= \kappa_\mathrm{dp} \Cp,
\end{align}
where $c_\mathrm{d}^{(i,j)}$ is the cost of computing the derivative of node $i$ with respect to its $j$-th child -- while the ratio $\kappa_\mathrm{dp}$ depends on the operations of the primal, usually, the derivative has a similar cost to the function itself (e.g. $\sin \mapsto \cos$), i.e. normally $\kappa_\mathrm{dp} \sim O(1)$.
The computation of the accumulated chained derivative costs one multiplication and addition per edge, i.e.
%
\begin{align}
    \Cc
    &= n_\mathrm{e} \cma \\
    &= \left(\sum_{i=1}^n l_\mathrm{node}^{(i)} + \sum_{i=1}^m l_\mathrm{param}^{(i)} \right) \cma \\
    &= (\bar l_\mathrm{node} n + \bar l_\mathrm{param} m) \cma \\
    &= \frac{(\bar l_\mathrm{node} n + \bar l_\mathrm{param} m) \cma}{\Cp}\Cp \\
    &= (\kappa_\mathrm{sp} + \kappa_\mathrm{mp}m)\Cp,
\end{align}
where $n_\mathrm{e}$ is the number of edges, $l_\mathrm{node}^{(i)}$ and $l_\mathrm{param}^{(i)}$ are the number of parent edges of the internal nodes and the parameter nodes, and the bar denoting average.
Adding up the three costs gives
%
\begin{equation}
    \frac{\Ca}{\Cp} = \frac{\Cp + \Cd + \Cc}{\Cp} = 1 + \kappa_\mathrm{dp} + \kappa_\mathrm{sp} + \kappa_\mathrm{mp}m,
    \label{eq:reverse_cost}
\end{equation}
which coincides with \cref{eq:adjoint_cost}.



\section{Accelerator physics vs machine learning}

\label{app:accphys_vs_ml}
A beamline can be seen as a neural network, where each beamline element is a network layer. Thus, it would seem that the remarkable efficiency of AD-REV in machine learning should transfer to beamline optimization. This is, alas, not necessarily the case.
Comparing networks, those of machine learning are shallow (even in deep learning) relative to those in beamlines. At the same time, they are wider and more connected. For example, while one dense neural network layer may add tens to thousands of parameters, a quadrupole only adds two. Similarly, the dense layer adds tens to thousands of nonlinearities, while a sextupole only adds two.\footnote{It is perhaps then not surprising that one of the few publications~\cite{Roussel2024} on application of AD to beamlines is on optimizing a neural network that generates the initial distribution, and not on optimizing the beam-line elements.}
For a beamline to have a significant number of parameters, such that using AD-REV is advantageous, it must be somewhat deep. Concomitantly, depth may be disadvantageous, as AD-REV requires keeping the graph of the operations of the primal in memory.
This suggests that a jump in application of AD-REV from beamline to lattice, wherein hundreds to thousands of turns are executed, may not be as straightforward as it might seem at first sight -- the depth of the DAG would be proportional to the lattice size and the number of turns.
In this case, potentially, the judicious choice might be AD-FWD, where the chained derivatives are computed and brought up along with the primal, foregoing the need to keep the graph in memory. Moreover, the results on GPU of \cref{sec:jutrack,sec:awake} suggest that the considerable theoretical disadvantage of forward mode with respect to reverse may not be as significant in practice and may therefore not be an impediment to the application of AD-FWD.

In essence, while we show theoretically and empirically that AD-REV (and AD-FWD) can and is usually be faster FD, it should not be assumed that, because it generally does so in other fields, it should do as well in accelerator physics -- while the methodology is similar, the specifics are not. 

\bibliography{apssamp}

\end{document}